\documentclass [12pt] {article}
\usepackage {amssymb, amsfonts}
\usepackage {amsmath}

\usepackage [cp866] {inputenc}
\inputencoding {cp866}
\begin {document}
\vskip0.5cm {\bf E.SH.GUTSHABASH} \vskip0.5cm

\vskip0.5cm \centerline {\bf {HYDRODYNAMICAL VORTEX ON THE PLANE}}

\vskip0.8cm \centerline {\bf {1. INTRODUCTION}} \vskip0.8cm
Hydrodynamics has always been a source of nonlinear equations
integrable by the inverse scattering method. To mention a few, the
Korteveg - de Vries, Kadomtsev - Petviashvili, Davey - Stewartson
and Bendjamin - Ono equations all have hydrodynamical origin. As a
rule, these equations can be obtained by direct application of
procedures of the multiscale decomposition type to the equations
of hydrodynamics. In other words, virtually all of them arise as a
result of certain reductions of the initial system of nonlinear
hydrodynamical equations, the latter having rather few known exact
solutions (one should mention here the Landau solution for the
submerged stream [1], and a rapidly developing approach to the
integration of nonlinear equations based on the group theory [2]).
It is natural then to search for completely integrable partial
cases of this system, since such cases, if found, would allow
regular procedures for constructing the solutions. This idea has
apparently first been realized in [3], where a Lax pair for the
Euler equation for ideal incompressible fluid was found. A more
special case - the vortex equation - was studied in [4].

In this paper we consider vortical solutions of the $2D$ Euler
equations, which have numerous applications, in particular, in
geophysical hydrodynamics [5,6] and plasma physics [7]. The
existence of a Lax pair (although, in a "weak" sense) for the
problem allows to apply a range of tools from the theory of
completely integrable systems.

Notice that the structure of the pair under consideration is
rather special - since both equations of the pair are of the first
order in the auxiliary function, no scattering problem occurs.
This leads, in particular, to solutions containing arbitrary
functions and certain other peculiarities.

The structure of the paper is as follows. In section 2 we give the
necessary preliminaries from the hydrodynamics and the theory of
vortices. Section 3 is devoted to solving a simplified vortex
equation, section 4 to constructing exact 2D solutions of the
vortex equation, both partial and more general, obtained by
application of the Darboux transformation.

\vskip0.8cm \centerline {\bf {2. PRELIMINARIES }} \vskip0.8cm

Let us consider the flow of an ideal incompressible
two-dimensional fluid in a domain $ {\mathbb D} \in {\mathbb R}^2
$. Let $ \vec v =\vec v (x, y)=(v_1, v_2)$ be the velocity field,
$p$ be the pressure, and $\rho $ be the density of the fluid. Then
the following system of equations on the velocity field holds [1]:

$$ {\mathrm div} \:\vec v=0, $$ $$ \eqno (2.1) $$ $$ \frac
{\partial {\vec v}}{\partial t}+(\vec v \nabla) \vec v=-\frac
{\nabla p} {\rho}. $$ The first equation in (2.1) expresses the
conservation of mass, the second one is the Euler equation.

This system should be provided with boundary conditions. It is
natural to assume that if the domain ${\mathbb D}$ is bounded,
then $$ (\vec v, \vec n)=0\:\:\:\mbox {at} \:\:\: (x, y) \in
{\partial {\mathbb D}}, \eqno (2.2) $$ where $\vec n$ is
orthogonal to the boundary surface, which means that the fluid can
not spill over the boundary, and that

$$ \vec v \to 0\:\:\:\mbox {at} \:\:\: \sqrt {x^2+y^2} \to \infty,
\eqno (2.3) $$ if the domain is unbounded.

As is well known, the system (2.1) have an infinite series of the
first integrals [8]. In particular, the quantities ( $k=1, 2.....
$)

$$I_k=\int \int_{\mathbb D} (\frac {\partial v_2}{\partial x}
-\frac {\partial v_1} {\partial y})^k dx\:dy \eqno (2.4) $$ are
first integrals.

Using the identity

$$ \frac {1} {2} \nabla v^2={\vec v}\wedge {\mathrm rot} \:
\vec v+(\vec v \nabla) {\vec v}, $$ we find from the second
equation in (2.1) that $(dw=dp/\rho)$

$$ \frac {\partial {\vec v}}{\partial t}-{\vec v} \wedge
{\mathrm rot} \:\vec v =-\nabla (w+\frac {v^2} {2}).$$ Applying
the rot operation to this equality, we obtain,

$$ \frac {\partial {\mathrm rot} \:\:\: {\vec v}} {\partial t} =
{\mathrm rot} \: [\:\vec v \wedge {\mathrm rot} \:\: {\vec v}].
\eqno (2.5) $$ This equation is called the vortex equation. It
shows, in particular, that in the stationary case the field of a
vortex commutes with the velocity field. Assuming, further, that
$v_1=-\Psi_y, \:v_2=\Psi_x$, where $\Psi=\Psi(x, y, t) $ is the
stream function {\footnote {Given the velocity field, it is always
possible to restore the stream function:

$$\Psi(x, y, t)=\int_{(x_0, y_0)}^{(x, y)} v_2 dx-v_1dy,
$$ where $(x_0, y_0) \in {\mathbb D}$ is a reference point, and
the value of the integral does not depend on the contour of
integration.}}, we find that ${\mathrm rot} \:\: {\vec v}={\vec k}
\triangle \Psi $, where
$\triangle=\partial^2_{xx}+\partial^2_{yy}$ is the Laplace
operator, ${\vec k}$ is the unit vector in the direction of $z$ -
axis. Setting $ \Omega =\triangle \Psi $ and defining the Jacobian
(or hydrodynamical bracket) by the relation

$$ J(w_1, w_2)=\frac {\partial (w_1, w_2)} {\partial x, \partial
y} \equiv \{w_1, w_2 \}_H=w_{1x} w_{2y}-w_{1y} w_{2x}, \eqno (2.6)
$$ we obtain the following nonlinear equation for vorticity $
\Omega $ {\footnote {it is easy to see that it can be re-written
in terms of the stream function alone.}}:

$$ \Omega_t=\{\Omega, \Psi \}_H.   \eqno (2.7) $$ It can be
checked by direct calculation that it is possible to re-write this
equation in the following hamiltonian form [9],

$$ \frac {\partial \Omega} {\partial t}=\{\Omega, {\cal H} \} _
{PB}, \:\:\:\:\: {\cal H} = \frac {1} {2} \int (\nabla \Psi)^2
dx\:dy, \eqno ( 2.8) $$ where ${\cal H}$ is the Hamiltonian, and
the Poisson bracket on smooth functionals is defined by

$$ \{F_1, G_1 \}_{PB} = \int \Omega\frac {\partial (\delta
F_1/\delta \Omega, \delta G_1/\delta \Omega)}{\partial (x, y)}
dx\:dy. \eqno ( 2.9) $$

To state the problem in a closed form, it is now necessary to
restore the stream function from vorticity. The corresponding
equation is reduced to an interior Dirichlet problem for the
Poisson equation ($F_0$ is a given function),

$$ \triangle \Psi =\Omega (x, y, t), \:\:\:\:\:\:\Psi_{|\partial
{\mathbb D}}=F_0 (l). \eqno (2.10) $$ Its solution has the form

$$\Psi (x, y, t)=\triangle^{-1} \Omega =\int \int_{\mathbb
D} G_0(x-x^{\prime}, y-y^{\prime}) \Omega (x^{\prime}, y^
{\prime}, t) dx^{\prime} dy^{\prime}+\int_{\partial {\mathbb D}}
\frac {\partial G_0} {\partial {\vec n}} F_0 (l) dl, \eqno (2.11)
$$ where $G_0(x, y)=(1/2\pi) \ln r+g_0(x, y) $ is the Green
function for the Dirichlet problem, $r=\sqrt {x^2+y^2}$, and $g_0
(x, y)$ is a harmonic function such that $G_0(x, y)_{|\partial
{\mathbb D}}=0 $. Notice that the function $\Psi$ in (2.10) is
defined up to addition of an arbitrary harmonic function
respecting the boundary condition, and satisfies certain
additional relation to be discussed below in detail.

To conclude this section, we single out two special cases of the
equation (2.7). The first (stationary case) is when the quantities
$ \Omega $ and $ \Psi $ are time - independent. We then have from
(2.7): $\Omega=F_0 (\Psi)$, where $F_0 (\Psi)$ is an arbitrary
function, and

$$ \triangle \Psi=F_0 (\Psi), \eqno (2.12) $$ that is, we obtain
an equation of elliptic type in dimension (2+0). Most interesting
are the completely integrable cases of the equation (2.12) with
functions $F_0 (\Psi) $ of the form $ \pm {e^{\pm \Psi}}, \:\pm
{\cosh \Psi}, \:\pm {\sinh \Psi}, \:\pm \sin \Psi, \:e^{\Psi}-e^
{-2\Psi}$ (elliptic version of the Tciceica's equation) and some
others, which allow to apply the inverse scattering method. A
range of such cases, with applications to various physical
phenomena, is considered in literature (see, for instance,
[10-12]). The methods used for solving (2.12) include the Hirota
method, Darboux and Backlund transformations [13-15], the
finite-gap integration [14], and the inverse scattering transform
both for the problem on the whole plane $(x, \:y)$ [16] and on the
half-plane $\{(x, \:y): y > 0 \} [17].$

Another important nontrivial partial case of the equation (2.7)
arises as follows. Assume that $\Omega=\triangle \Psi=F_1 (\Psi)
+g $ with a certain function $F_1$, and $g=g(x, y, t)$. Assume
further that either $F_{1\Psi}=0$, or $\Psi_t=0$. We then obtain
from (2.7),

$$g_t=\Psi_yg_x-\Psi_xg_y.  \eqno (2.13) $$. This equation, which
is given, for instance, in [18], we shall call the simplified
vortex equation. As well as (2.7), this equation is of significant
physical interest, and, importantly, is completely integrable.

\vskip1.5cm \centerline {\bf 3. EXACT SOLUTIONS OF EQUATION
(2.13).}

\vskip0.8cm

In this section we construct an exact solution of the simplified
vortex equation (2.13) assuming that $g =\triangle \Psi+f(x, y,
t)$, where $f(x, y, t)$ is a known function.

It is easy to verify by straightforward computation that the
equation (2.13) can be represented as a compatibility condition
for the following system of scalar equations {\footnote {In the
present form this Lax pair was suggested by A.Ja.Kazakov.}}:

$$ (
g-\lambda)\varphi=0,\:\:\:\:\varphi_t=\Psi_y\varphi_x-\Psi_x\varphi_y,
\eqno (3.1) $$ where $\varphi=\varphi (x, y, t)$ is a
complex-valued function and $\lambda \in {\mathbb C}$ is the
spectral parameter. It is convenient for our purposes to put the
first equation of the pair in a somewhat different form to obtain,

$$
 g_x\varphi_y=g_y\varphi_x,\:\:\:\:\varphi_t=\Psi_y\varphi_x
-\Psi_x\varphi_y. \eqno (3.2) $$ The existence of such
representation allows us to construct a family of exact solutions
of the equation (2.13). To this end, we use the method of the
generalized Darboux transformation {\footnote {This method is also
known as Moutard's transform.}} (see, for example, [19]).

Let

$$ \tilde {\varphi} (x, y, t) = \frac {Q (\varphi, \varphi_1)}
{\varphi_1 (x, y, t)}, \:\:\:\:\:\:\:\:\:Q (\varphi, \varphi_1) =
\int_{(x_0, y_0, t)}^{(x, y, t)} q (\varphi, \varphi_1), \eqno
(3.3) $$ where $q(\varphi,\varphi_1)=
(\varphi_x\varphi_1-\varphi\varphi_{1x})dx+(\varphi_y\varphi_1-
\varphi_{1y} \varphi) dy$ is a 1-form, and $ \:\varphi_1 $ is a
fixed solution of (3.2); then the integral $Q (\varphi, \varphi_1)
$ does not depend on the contour of integration by virtue of the
first equality. Let us verify the covarianse of the system (3.2)
under the transformation $ \varphi \to {\tilde \varphi}, \:g \to
{\tilde g}, \:\Psi \to {\tilde \Psi} $. For the first equation
this leads to a dressing relation of the form

$$ {\tilde g}_y={\tilde g}_x\frac {\varphi_{1y}} {\varphi_
{1x}}, \eqno (3.4) $$ which implies that{\footnote {For a first
order partial differential equation of the form $\{Y, A \}_H=0$,
with a given $A=A(x, y)$, the function $Y=R_0(A(x, y))$ is a
solution for any function $R_0 (.)$.}}

$$ {\tilde g} (x, y, t)=f_0 (\varphi_1 (x, y, t)), \eqno (3.5)
$$ where $f_0 (\varphi_1 (...))$ is an arbitrary function. The
verification of the second relation gives

$$ Q_t-Q_x{\tilde \Psi}_y+Q_y {\tilde \Psi}_x+Q [-(\ln
\varphi_1)_t+{\tilde \Psi}_y (\ln \varphi_1)_x-{\bar \Psi}_x (\ln
\varphi_1)_y] =0. \eqno (3.6) $$ Solving this linear first order
partial differential equation (recall that the function $\tilde
\Psi$ is known), we find $Q$ explicitly.

Taking (3.5) into account, we have from (2.11),

$$
{\tilde \Psi} (x, y, t)=\int \int_D G_0 (x-x^{\prime},
 y-y^{\prime},t)[f_0(\varphi_1(x^{\prime},y^{\prime},t))-f(x^{\prime},
y^{\prime}, t)] dx^{\prime} dy^{\prime}, \eqno (3.7) $$ which
solves the initial problem.

Clearly, the procedure suggested can be repeated and one can
construct the $N$-dressed solutions, but we shall not dwell on
that in detail here.

 \vskip1.5cm \centerline {\bf 4. SOLUTION OF THE VORTEX EQUATION.}
 \vskip0.8cm

1. {\em Linearized equation.}

Let us first consider the linearization of the equation (2.7). Let
$ \Psi (x, y, t)=\Psi_1(x, y, t)+\Psi_0 (x, y, t),\:\Omega (x, y,
t) = \Omega_1(x, y, t)+\Omega_0 (x , y, t) $, where $ \Omega_0
=\triangle \Psi_0=0, \:\Psi_0$ is a known solution of (2.7). We
then have

$$ (\triangle \Psi_1)_t=(\triangle \Psi_1)_x\Psi_{0y}-
(\triangle \Psi_1)_y\Psi_{0x}. \eqno (4.1) $$ Taking $\triangle
\Psi_1$ in the form

$$ \triangle \Psi_1 \sim a (\hat {\alpha}, \:\hat {\beta}, \:\hat
{\omega}) \exp \{i ({\hat \alpha} x + {\hat \beta} y + {\hat
{\omega}} t) \}, \:\:\: \mbox {where} \:\:\: a (\hat {\alpha},
\:\hat {\beta}, \:\hat {\omega}) = {\bar a} (-\hat {\alpha}, \:-
{\hat {\beta}}, \:- {\hat {\omega}}),  \eqno (4.2) $$ and
substituting it in (2.7), we obtain the following dispersion
relation for the waves (assuming that $ \Psi_{0x}=p_0, \:\Psi _
{0y}=q_0 $, where $ \: p_0, \:q_0 $ are constants)

$$
{\hat \omega}=p_0{\hat \alpha}-q_0 {\hat \beta}. \eqno (4.3)
$$

2. {\em Certain partial solutions.}

Using various approaches, we now construct a set of partial
solutions of (2.7).

a). Let $\Psi(x, y, t)=(X(x)+Y(y))T(t)$. We then have from (2.7):

$$ a(X_{1x}+Y_{1y})=X_{1xx}Y_1-Y_{1yy}X_1,$$ where
$X_1=X_x, \:Y_1=Y_y, \:a=T_t/t^2={\mathrm const}.$ Searching for
the functions $X_1, \:Y_1$ polynomial in $x$ and $y$ respectively,
we find,

$$ \Psi(x, \:y, \:t)=\bigl [c (x^2-y^2)+d_1x+d_2y+h \bigr]
\frac {a} {t}, \:\:\:\Omega =\triangle \Psi=0, \eqno (4.4) $$
where $t > 0$, and $ \:c, \:d_1, \:d_2, \:h $ are constants. This
solution is quadratic in the spatial variables and slowly (sub-
exponentially) decreases in time. Obviously, it describes a
vortex-free flow of the fluid.

b). Let $ \Psi=X (x)+Y(y)+T (x, y, t) $. We then have,

$$ (
T_{xx}+T_{yy})_t=(X_{xx}+T_{xx}+T_{yy})_x(Y_y+T_y)-(Y_{yy}+T_{xx}+T_{yy})_y
(X_x+T_x) . $$ Letting
$X=a_1x^2+b_1x+c_1,\:\:Y=a_2y^2+b_2y+c_2,\:T=c\cos(\alpha x+\beta
y +\gamma t+\delta) $, where $a_i, \:b_i, \:c_i,
 \;i=1,2,\:c,\:\:\alpha,\:\beta,\:\gamma,\:\delta$ are constants, we obtain: $a_1=a_2=0 $,

$$ \Psi (x, y, t) =b_1x+b_2y+d+c\cos (\alpha x+\beta y+\gamma t
+\delta), $$ $$ \eqno (4.5) $$ $$ \Omega =-c (\alpha^2 +\beta^2)
\cos (\alpha x+\beta y+\gamma t+\delta), $$ and $ \gamma =\alpha
b_2-\beta b_1$. These formulae were used, in particular, in [20]
to construct solutions on the basis of group-theoretical
considerations and describe a stream function linear in spatial
variables on the background of spatial-temporal oscillations.

In a similar way, one can obtain a solution growing in the
absolute value in all variables:

$$
\Psi (x, y, t)=b_1x+b_2y+d+c\cosh (\alpha x+\beta y+\gamma t
+\delta),
$$
$$
 \eqno (4.6)
$$ $$ \Omega=c (\alpha^2+\beta^2) \cosh (\alpha x+\beta y+\gamma
t+\delta), $$ where $ \gamma=\alpha b_2-\beta b_1.$

c). Let us define the complex variables $z=x+iy, \: {\bar z} =x-iy
$, and let $\partial=(1/2)(\partial_x-i\partial_y),\: \bar {
\partial}=(1/2)(\partial_x+i\partial_y).$ On passing to these variables
in (2.7), we obtain

$$ \Psi_{z {\bar z} t} =2i(\Psi_{z {\bar z} {\bar z}}
\Psi_z-\Psi_{zz {\bar z}} \Psi_{\bar z}). \eqno (4.7) $$ Now,
setting $ \chi =\Psi_{\bar z}$, we have,

$$ \chi_{zt} =2i(\chi_{z {\bar z}} {\bar \chi}-\chi_{zz}
\chi). \eqno (4.8) $$ Assume that $\chi_{\bar z}=0$, that is, $
\chi $ is an analytic function in the domain $ {\mathbb D} $. Then
it satisfies the following nonlinear equation,

$$\chi_{zt}=-2i\chi_{zz} \chi. \eqno (4.9) $$ In the
simplest case when $ \chi (z, \:t)=Z(z)T(t)$, this gives,

$$
\frac {T_t}{T^2} \equiv a^2, \:\:\:\:
 {\mathrm Ei} \: (\ln Z)={\mathrm Ei} \: (\ln
Z_0)+\frac {a^2} {2i} e^{\frac {C_1} {a^2}} (z-z_0), \eqno (4.10)
$$ where $ {\mathrm Ei} \: (u)=\int_{-\infty}^u (e^t/t) dt $
is the integral exponential function, $a, \:C_1 $ are arbitrary
constants, and $Z_0=Z(z_0)$. Formally, one can write,

$$
Z(z)=e^{{\mathrm Ei}^{-1} [\ln
 Z_0+\frac{a^2}{2i}e^{\frac{C_1}{a^2}}(z-z_0)]}.\eqno(4.11)
$$

Another solution of (4.9) is found if we plug $ \chi $ in the form
$ \chi (z, t)=\chi (\xi) $, where $ \xi=\alpha z+\beta t+\gamma $,
$ \alpha, \:\beta, \:\gamma $ being constants. The solution is
then given by

$$ \chi (z, t)=\xi. \eqno (4.12) $$ In both cases (4.10) -
(4.11) and (4.12), using the Green-Cauchy representation (the $
\bar \partial $ - problem relation) [21], we obtain for the
function $\Psi$ in the case of a bounded domain $\mathbb D$,

$$ \Psi(z, {\bar z}, t)=\frac {1} {2\pi i}\int_{\partial
{\mathbb D}} \frac {\chi_0 (\zeta, {\bar \zeta}, t)} {\zeta-z}
d\zeta +\frac {1} {2\pi i}\int \int_{\mathbb D} \frac {\chi
(\zeta, {\bar \zeta}, t)}{\zeta-z} d\zeta\wedge d {\bar \zeta},
\eqno (4.13) $$ where the domain $ {\mathbb D} $ is assumed
bounded, and $ \partial {\mathbb D} $ stands for its boundary,
$\chi_0(z, {\bar z}, t)=\Psi(z, {\bar z}, t)_{|z \in \partial
{\mathbb D}}$.

2. {\em Lax representation and the method of Darboux
transformation.}

The equation (2.7) can be represented as a compatibility condition
for the following over-determined system of scalar equations [4,
22] {\footnote {From the viewpoint of the spectral theory the
first equation in (4.14) can be interpreted as an eigenvalue
problem for the operator of "generalized" \enskip rotation of the
plane, $\Omega_x\frac {\partial} {\partial y} -\Omega_y\frac
{\partial} {\partial x}$, which is easily shown (see, for
instance, [2]) to be antisymmetric.}}:

$$
 L\varphi=\lambda\varphi,\:\:\:\:\varphi_t=-A\varphi, \eqno (4.14)
$$ where $L\varphi=\{\Omega, \varphi \} _H, \:A\varphi = \{\Psi,
\varphi \} _H, \:\lambda \in {\mathbb C} $ is the spectral
parameter, and the symbol $ \{.. \}_H $ refers to the
"hydrodynamical"\enskip bracket defined by (2.6). Here the
compatibility, and hence the integrability of (2.7), is understood
in a "weak" \enskip sense, that is, under the additional condition

$$
\{\Omega_t+\{\Psi,\Omega \}_H,\varphi \}_H=0.   \eqno(4.15)
$$

We now apply the method of Darboux transformation to construct a
solution of the equation (2.7). Two different versions of the
method will be considered.

i). We use a matrix transform. It is easy to notice that the
system (4.14) can be rewritten in a matrix form (taking into
account that $ \Psi = {\bar \Psi}, \:\Omega = {\bar \Omega} $),

$$
 \Omega_x\Phi_y-\Omega_y\Phi_x=\Phi\Lambda,\:\:\:\:\:\:\Phi_t=\Psi_y\Phi_x-
\Psi_x\Phi_y, \eqno (4.16)
$$
where

$$
\Phi=\Phi (x, y, t, \lambda)=\left (\begin {array} {cc} 0&\varphi\\
                         {\bar \varphi}&0\end{array}\right),\:\:\:\:\:\:
\Lambda={\mathrm diag} ({\bar \lambda}, \:\lambda), $$ and the
condition (4.15) is assumed to hold.

Let us check the covariance of the equations (4.16), $ \Phi \to
{\tilde \Phi}, \:\Psi \to {\tilde \Psi}, \:\Omega \to {\tilde
\Omega} $, under the transformation

$$
{ \tilde \Phi} = \Phi-\sigma_1 \Phi
 \Lambda,\:\:\:\sigma_1=\Phi_1\Lambda_1\Phi_1^{-1}, \eqno (4.17)
$$ where $ \Phi_1 =\Phi (x, y, t, \lambda_1) $ is a solution of
the system (4.16) corresponding to a fixed value $ \lambda
=\lambda_1$. We then arrive at the following "dressing" relations,

$$ {\tilde \Psi}_x=\Psi_x, \:\: {\tilde \Psi}_y=\Psi_y, \:\:
\{{\tilde \Omega_x}, \Omega \}_H=0, \:\: \{{\tilde \Omega},
\sigma_1 \}_H=0, \:\: {\tilde \Omega}_x=\Omega_x,  \eqno (4.18)
$$ from which we infer that

$$ {\tilde \Psi}=\Psi+F_1, \:\:\: {\tilde \Omega}=\Omega+G_1,
\eqno (4.19) $$ where $F_1=F_1 (t), \:G_1=G_1 (t)$ are arbitrary
functions of time.

Notice, further, that the substitution $ \Psi \to \Psi+F_1 (t)+h
(x, y, t), \:\Omega \to \Omega+G_1 (t) $,

$$ h=h (x, y, t)=\sum_{i=1}^L d_i (t) h_i (x, y), \eqno
(4.20) $$  where the integer $L$ and the functions $d_i (t),
\:i=1, \:L$ are arbitrary, and $ h_i (x, y) $ are harmonic, does
not change the form of the equation (2.7) if the condition
\footnote {This remark can be generalized to the case $h (x, y, t)
=\sum_{i=1}^L h_i (x, y, t), \:\triangle h=\triangle h_i=0, \:i=1,
\:L $.}

$$ G_{1t}=\{\Omega, \: \sum_{i=1}^L d_i(t) h_i \}_H, \eqno
(4.21) $$ is satisfied.

This allows to write down the new solution of the equation (2.7)
as follows,

$$
{\tilde \Psi}(x, y, t)=\Psi (x, y, t)+F_1 (t)+h (x, y, t) .\eqno
(4.22)
$$

ii). We now provide an alternative Darboux transformation, which
apparently makes it possible to construct a wider class of
solutions, as compared to the above. To this end, we check the
co-invariance of the system (4.14) under the transformation $
\varphi \to {\tilde \varphi}, \:\Psi \to {\tilde \Psi}, \:\Omega
\to {\tilde \Omega} $, where

$$ {\tilde \varphi}=T(x, y, t)(\varphi_x-\sigma_1\varphi),
\eqno (4.23) $$ $ \sigma_1=(\ln \varphi_1)_x, $ $ \varphi_1 $ is a
fixed solution of (4.14) corresponding to $\lambda =\lambda_1 $,
and $T=T (x, y, t)$ is a function to be determined later
{\footnote {In paper [22] the transformation (4.23) was used in
the case $ \lambda=0 $, $T (x, y, t)=1/\Omega_x. $}. Matching the
coefficients at $ \varphi, \:\varphi_x $ and $\varphi_{xx}$, we
then obtain, after a cumbersome calculation, the following
over-determined system of dressing relations,

$$
\{{\tilde \Omega}, \:\Omega \}_H=0, \:\:\:\: \{{\tilde \Psi}
-\Psi, \: \Omega \}_H=0, \eqno (4.24a)
$$
$$
\lambda T [\sigma_1-\frac {{\tilde
 \Omega}_x}{\Omega_x}(\frac{\Omega_{xx}}{\Omega_x}+\sigma_1)]-\{{\tilde
\Omega}, T\sigma_1 \}_H =0, \eqno (4.24b)
$$
$$
\{{\tilde \Omega}, \ln T \}_H +{\tilde \Omega}_x [(\frac
{\Omega_y} {\Omega_x})_x-
 \sigma_1\frac{\Omega_y}{\Omega_x}]+{\tilde
\Omega}_y\sigma_1 +\lambda (\frac {{\tilde \Omega} _x}{\Omega_x}
-1)=0, \eqno (4.24c)
$$
$$
(T\sigma_1)_t+\{{\tilde \Psi}, T\sigma_1 \}_H-\lambda T [\frac
{\Psi_x-{\tilde \Psi}_x}
{\Omega_x^2}(\Omega_x\sigma_1+\Omega_{xx})-\frac {{\Psi}_{xx}}
{\Omega_x}]=0, \eqno (4.24d)
$$
$$
( \ln T) _t+{\tilde \Psi}_x [(\ln T)_y+(\frac {\Omega_y}
{\Omega_x})_x-\sigma_1 \frac {\Omega_y} {\Omega_x}]+{\tilde \Psi}
_y [- (\ln T)_x+\sigma_1]+
$$
$$
\eqno (4.24e)
$$
$$
 +[\Psi_{yx}-\Psi_x(\frac{\Omega_y}{\Omega_x})_x-\Psi_{xx}
\frac {\Omega_y}{\Omega_x}-\frac {\sigma_1}{\Omega_x}\{\Omega,
\Psi \}_H]+\frac {\lambda}{\Omega_x}(\tilde \Psi_x-\Psi_x)=0.
$$

Let us analyze the compatibility of this system, limiting
ourselves to simpler reductions. Obviously, there are two
essentially different cases: $ \lambda=0 $ and $\lambda \ne 0 $,
when the coefficients at $\lambda$ should vanish.

Let $\lambda \ne 0$. We then infer that ${\tilde \Psi}_x =\Psi_x$
from (4.24e); $\Psi_{xx}=0 $ from (4.24d); $ {\tilde \Omega}
_x/\Omega_x=1$ from (4.24c). On taking into account (4.24a), we
obtain that

$$ {\tilde \Psi}=[\alpha_0 (y, t)+c_1]x+c_2, $$ where $
\alpha_0 (,)$ is an arbitrary function, and $c_1, \:c_2 $ are
arbitrary constants. From this we find that $ {\tilde \Omega}=
\alpha_{0yy} x, \: {\tilde \Omega}_y=\alpha_{0yyy} x, \: {\tilde
\Omega}_x =\alpha_{0yy} $, and that $\Omega_y/\Omega_x=(\ln \alpha
_{0yy})_yx$ in view of the first equality in (4.2a). Eventually,
we arrive at a rather complicated over-determined system of
equations on the function $\alpha_0$, which is likely to be
incompatible.

Now let $\lambda=0, \:T=1$. On calculating, we then obtain that
(4.24) transforms into a system of the form

$$\{{\tilde \Omega}, \Omega \}_H=0, \:\: \{{\tilde \Omega},
\sigma_1 \}_H =0, \:\: \{{\tilde \Psi}-\Psi, \Omega \}_H=0, \:\:
\{{\tilde \Omega}, \Psi_x \}_H=0, $$ $$ \eqno (4.25) $$ $$
(\frac{\Omega_y}{\Omega_x})_x=0,\:\:\:\sigma_{1t}+\{{\tilde \Psi},
\sigma_1 \}_H=0. $$ Substituting

$${\tilde \Psi}=\Psi+U_1, \:\: {\tilde \Omega}=\Omega+V_1,
\:\:\:V_1 =\triangle U_1, \eqno (4.26) $$ in (4.25), we get the
following system of equations on $U_1 $:

$$ \{\Omega+V_1, \Omega \} _H=0, \:\: \{\Omega+V_1, \sigma_1 \}
_H=0, \:\: \{\Omega+V_1, \Psi_x+U_{1x} \}_H=0, $$ $$ \eqno (4.27)
$$ $$\{U_1, \Omega \}_H=0, \:\:\:\sigma_{1t}+\{\Psi+U_1, \sigma_1
\}_H=0. $$ The compatibility of the first three equations becomes
obvious if we set $\Omega+V_1=p_0(\Omega, \sigma_1,
\Psi_x+U_{1x})$, where $p_0(.)$ is an arbitrary function. It then
follows from the first relation that $\{\triangle U_1, \Omega
\}_H=0$, that is, $\triangle U_1=q_0(\Omega)$ where $q_0(.)$ is an
arbitrary function, hence $U_1 =\triangle^{-1}q_0 (\Omega)\equiv
g_0(\Omega)$. This means that the fourth equation in (4.27) is
compatible with the first three. Moreover, the fourth equation in
(4.27) implies that $U_{1y} /U_{1x}=\Omega_y/\Omega_x$, and the
requirement of the compatibility with the remaining equation gives

$$U_1(x, y, t, \lambda_1)=-\int_{-\infty}^x \frac {\sigma _
{1t}+\{\Psi, \sigma_1 \}_H}{\{\Omega, \sigma_1 \}_H} dx \eqno
(4.28) $$ under the assumption that $U_1=U_1 (x, y, t, \lambda_1),
\:U_1(-\infty, y, t, \lambda_1)=0$ and $\{\Omega, \sigma_1 \}_H
\ne 0 $.

Thus, we have obtained dressing relations allowing to construct a
new explicit solution of the equation (2.7) from a known one. It
is also possible to obtain the corresponding multiple dressing
formulae in the spirit of the Darboux transform method [23].

\vskip1.5cm \centerline {\bf 5. CONCLUSION} \vskip0.8cm

Several procedures for constructing the solutions of the $2D$
vortex equation have been suggested in this paper. A number of
questions remains open, however: the meaning of the notion of "
weak integrability", the possibility of application of the
multidimensional inverse scattering transform method, the
Hamiltonian formalism, and, especially, quantization of the
vortices, to mention a few.

Notice that various extended versions of this equation are of
doubtless interest. For instance, the equation for the so-called
Rossby waves [6, 7] has the form

$$ \Omega_t=\{\Omega, \Psi \}_H-\beta\Omega, $$ where $\beta$ is
a parameter, that is, it differs from (2.7) by an additional term.
As shown in [4], it also admits a Lax representation (which is
easily recovered from (4.14)), hence exact solutions can be
constructed.

Let us emphasize that there is a wide range of adjacent topics
related to the theme of this paper: the problem of decay of a
turbulent spell, models of vortex chains, behaviour of vortex
ensembles in geophysics, plasma physics, and many others. These
reasons urge a further analysis of the equation (2.7) and its
generalizations.

The author is grateful to A.V.Urov and P.P.Kulish for support.

\vskip2.0cm

1. \parbox [t] {12.7cm}
    {{\em L.D.Landau, E.M.Lifshits.} Hydrodynamics. M., Science (1997).}

 \vskip0.3cm
2. \parbox [t] {12.7cm}
     {{\em P.Olver.} Applications of Lie groups to differential equations.
     Springer-Verlag (1980).}

\vskip0.3cm 3. \parbox [t] {12.7cm}
    {{\em S.Friedlander, M.Vishik.} Phys. Lett. {\bf 148A}, 6, 7 (1990).}

 \vskip0.3cm
4. \parbox [t] {12.7cm}
     {{\em Y.Li.} J.Math. Phys. {\bf 41}, 728
      (2000).}

\vskip0.3cm 5. \parbox [t] {12.7cm}
     {{\em J. Pedloski.} Geophysical hydrodynamics. In 2 vols. M., World,
     (1984).}

\vskip0.3cm 6. \parbox [t] {12.7cm}
     {{\em A.C.Monin.} Theoretical bases of geophysical
     hydrodynamics. Leningrad, Gidrometeoizdat, (1988).}

\vskip0.3cm 7. \parbox [t] {12.7cm}
     {{\em V.M.Chernousenko, V.M.Kuklin, I.P.Panchenko.} In:
     "Integrability and kinetic equations for solitons ". Naukova
     dumka, Kiev, (1990).}

\vskip0.3cm 8. \parbox [t] {12.7cm}
     {{\em V.I.Arnold.} Mathematical methods of the classical mechanics.
     M., Science, (1979).}

\vskip0.3cm 9. \parbox [t] {12.7cm}
     {{\em V.E.Zakharov, E.A.Kuznetsov.} UFN {\bf 167}, 1137 (1999).}

\vskip0.3cm 10. \parbox [t] {12.7cm}
    {{\em O.B.Kaptsov.} JETP {\bf 98}, 532 (1990).}

 \vskip0.3cm
11. \parbox [t] {12.7cm}
    {{\em O. Hudak.} Phys. Lett. {\bf 89A}, 245 (1982).}

\vskip0.3cm 12. \parbox [t] {12.7cm}
    {{\em Sh. Takeno.} Prog. Theor. Phys. {\bf 68}, 992 (1982).}

\vskip0.3cm 13. \parbox [t] {12.7cm}
    {{\em A.B.Borisov, G.G.Taluts and others.} In " Modern problems
    of the theories of magnetism". Kiev, Naukova dumka (1986).}

\vskip0.3cm 14. \parbox [t] {12.7cm}
    {{\em M.B.Babitch, E.Sh.Gutshabash and others.} In " High-temperature
    Superconductivity. Urgent problems". Leningrad, LGU (1991).}

\vskip0.3cm 15. \parbox [t] {12.7cm}
    {{\em G.Leibbrandt.} Phys. Rev. {\bf 15B}, 3351 (1977).}

\vskip0.3cm 16. \parbox [t] {12.7cm}
    {{\em A.B.Borisov, V.V.Kiseliev.} Inv. Prob., {\bf 5}, 959 (1989).}

\vskip0.3cm 17. \parbox [t] {12.7cm}
    {{\em E.Sh.Gutshabash, V.D.Lipovski, S.S.Nikulichev.} TMP {\bf 115},
    323 (1998).}

\vskip0.3cm 18. \parbox [t] {12.cm}
     {{\em M.A.Salle.} Private communication (1992).}

\vskip0.3cm 19. \parbox [t] {12.7cm}
    {{\em E.Sh.Gutshabash.} JETP Letters {\bf 78}, 740 (2003).}

\vskip0.3cm 20. \parbox [t] {12.7cm}
    {{\em S.Y.Lou, X.Y.Tang and others.} nlin. PS/0509039.}

\vskip0.3cm
 21. \parbox [t] {12.7cm}
    {{\em I.N.Vekua.} Generalized analytical functions. M., Science (1988).}

\vskip0.3cm 22. \parbox [t] {12.7cm}
    {{\em Y.Li, A.V.Yurov.} Stud. Appl. Math. {\bf 111}, 101 (2003).}

\vskip0.3cm 23. \parbox [t] {12.7cm}
    {{\em V.B.Matveev, M.A. Salle.} Darboux Transformation and
    Solitons. Springer-Verlag. (1991).}

\vskip1.2cm
 \centerline {Institute Research for Physics St.-Petersburg State
University,} \centerline {St.-Petersburg, Staryi Peterhoff,
198504,} \centerline {e-mail: gutshab@EG2097.spb.edu}

 \vskip0.4cm

\end {document}